\documentclass[3p,numbers,sort&compress]{elsarticle}

\usepackage{amsmath,amssymb,amsthm}
\usepackage[english]{babel}
\usepackage{graphicx, color, epsfig}
\usepackage{enumerate}
\usepackage{latexsym}
\usepackage{fancyhdr}
\usepackage{multirow}

\newcommand{\ds}{\displaystyle}
\newcommand{\vs}{\vspace}
\newcommand{\nt}{\nonumber}

\newcommand{\A}{\mathbf{A}}

\newcommand{\B}{\mathbf{B}}
\newcommand{\C}{\mathrm{C}}

\newcommand{\dd}{\mathrm{d}}

\newcommand{\E}{\mathbf{E}}
\newcommand{\e}{\mathrm{e}}
\newcommand{\ee}{\mathrm{e}}

\newcommand{\g}{\mathrm{g}}
\newcommand{\G}{\mathrm{G}}
\newcommand{\h}{\mathrm{h}}

\newcommand{\I}{\mathrm{I}}

\renewcommand{\j}{\mathrm{j}}

\newcommand{\kk}{\mathbf{k}}

\newcommand{\nn}{\mathbf{n}}
\newcommand{\mm}{\mathbf{m}}
\newcommand{\pp}{\mathbf{p}}

\newcommand{\rr}{\mathbf{r}}

\newcommand{\Y}{\mathrm{Y}}

\newcommand{\sigmag}{\textrm{\mathversion{bold}$\sigma$}}
\newcommand{\zerog}{\textrm{\mathversion{bold}$0$}}

\newcommand{\ldc}{[\![}
\newcommand{\rdc}{]\!]}
\newcommand{\pr}{\partial}

\newcommand{\lambdabarre}{\!\!\!
\raisebox{0.04cm}{$\begin{array}{c}
-
\\[-0.35cm]
\lambda
  \end{array}$}\!\!\!}

\newcommand{\lnd}{~\!\!\!\!\!\!&}
\newcommand{\rnd}{&\!\!\!\!\!\!~}

\newcommand{\Si}{\textrm{Si}}

\begin{document}
  \begin{frontmatter}
\title{Lamb shift of interactive electron-hole pairs in spherical semiconductor quantum dots}

\author[UCP]{B. {\sc Billaud}\corref{corresponding_author}}
\ead{bbillaud@u-cergy.fr}
\author[UCP]{T.-T. {\sc Truong}}
\cortext[corresponding_author]{To whom correspondence should be addressed.}

\address[UCP]{Laboratoire de Physique Th\'eorique et Mod\'elisation,
\\
Universit\'e de Cergy-Pontoise/CNRS UMR 8089,
\\
F-95302 Cergy-Pontoise Cedex, France.}
    \begin{abstract}
The ground state Lamb shift of a semiconductor spherical Quantum Dot is computed in the effective mass approximation. It appears to be significant enough to be detectable for a wide range of small Quantum Dots synthesized in semiconductors. A possible way to observe it, via the Casimir effect, is suggested.
    \end{abstract}
  \end{frontmatter}
\section{Introduction}
The investigation of atomic physics properties of semiconducting Quantum Dots (QDs) is a popular topic for fundamental as well as applied physics. Such nanostructures restrict the motion of charge carriers to a confined region of space. Two QDs are never identical, because of the crucial role of phonons, surface effects or bulk disorder on their electronic structure, but they may be advantageously considered as artificial giant atoms. Thanks to their adjustable quantized energy spectrum, controlled only by their size, they may be used, for example, as semiconductor lasers \cite{Kirkstaedter_1994}, single-photon sources \cite{Yoshie_2004}, {\it qubits} \cite{Trauzettel_2007}, single-electron transistors \cite{Ishibashi_2003}, artificial {\it fluorophores} \cite{So_2006}. In the early 1980s, the so-called Quantum Size Effects (QSE), characterized by a blue-shift of their optical spectra, has been observed in a large range of strongly confined systems \cite{Vojak_1980, Ekimov_1980, Golubkov_1981, Kash_1986, Temklin_1987}. It comes from a widening of the semiconductor optical band gap, due to the increase of the charge carriers confinement energy \cite{Yoffe_2002}. Modern methods as well as empirical pseudo-potential methods or {\it ab initio} approaches, such as the Density Functional Theory (DFT), which are appropriate for numerical determination of semiconductor bulk band structures or confinement energies, are discussed in Refs. \cite{Delerue, Bester_2009}. However, to the best of our knowledge, no comprehensive approach, which offers a significant analytic treatment, seems to exist at present.

In this work, to explore their analogy with real atoms, we look at the interaction of a spherical semiconductor QD with a quantized electromagnetic field and investigate the Lamb shift of its energy levels. Discovered and actively studied in the late 1940s \cite{Lamb_1947, Bethe_1947, Welton_1948, Bethe_1950}, the Lamb shift still remains nowadays an area of intense research, for example within dielectric materials \cite{Sajeev_1990}, on the excitation modes of an electromagnetic cavity \cite{Brune_1994}, or on the coupling of electromagnetic modes to the semiconductor QD surface \cite{Rotkin_2000}. Recent works also present experimental protocols, in which Lamb shifts enhanced by an electromagnetic cavity are measured in nanostructures like {\it transmon} \cite{Fragner_2008} or studied in semiconducting QDs coupled to a planar slab of negative-index material \cite{Yao_2009}. However, the Lamb shift, well understood for real atoms, seems not to be approached neither theoretically nor experimentally in semiconductor QDs.

The object of this paper is to fill this gap by offering a consistent theoretical treatment leading to the evaluation of the shifts for a wide range of realistic spherical semiconductor QDs. To describe a confined electron-hole pair, it seems appropriate to use the popular effective mass approximation (EMA) model \cite{Brus_1984, Efros_1982, Kayanuma_1988, Billaud_2009a}. We first review the tools needed to work out the Lamb shift in section {\bf\ref{sec_2}}, and the basic properties of a spherical semiconductor QD in section {\bf\ref{sec_3}}. From there on, we deduce, in sections {\bf\ref{sec_4}} and {\bf\ref{sec_5}}, the Lamb shift for a massive charged particle confined by an infinite potential well. As this Lamb shift is {\it negative}, we comment, in section {\bf\ref{sec_6}}, our result to emphasize the difference with the positive Lamb shift observed in real atoms. In section {\bf\ref{sec_7}}, for spherical semiconductor QDs, we show that there exists a possibility to observe experimentally the Lamb shift, at least in a so-called strong confinement regime and for a judiciously chosen semiconductor. We also suggest a protocol, built on the Casimir effect, which might allow to observe this predicted Lamb shift, despite the fundamental non-degeneracy of semiconducting QD levels. A concluding section summarizes our main results and indicates possible future research perspectives.
\section{Considerations on Lamb shift} \label{sec_2}
In 1947, investigations on Zeeman effect in hydrogen atom showed that its fine structure does not agree with the predictions of Dirac theory \cite{Lamb_1947}. The $2s$-level, which should be degenerated with $2p$-levels, is actually shifted by an energy $\approx$1057MHz, the so-called Lamb shift. Theoretical work demonstrated that it arises from the coupling of the electron motion with the surrounding quantized electromagnetic field \cite{Bethe_1947, Welton_1948}. Consider a non-relativistic spinless massive particle of mass $m^*$ and of charge $qe$ interacting with a quantized electromagnetic field\footnote{Relativistic corrections can generally be calculated \cite{Feynman_1948, Schwinger_1949}, but they are not relevant in the comprehension of Lamb effect, particularly in spherical semiconductor QDs, where confined charged carriers are considered as non-relativistic particles.}. It is represented by the Pauli-Fierz Hamiltonian in the Coulomb gauge \cite{Pauli_1938}, here written in units where $\hbar=c=1$,
  $$
H_\mathrm{PF}=H_0+eH_\mathrm{int}+e^2H'_\mathrm{int}+H_\mathrm{em}.
  $$
The particle Hamiltonian in absence of electromagnetic field $H_0=\frac{\pp^2}{2m^*}+V(\rr)$ is supposed to be diagonal in an orthonormal basis of its eigenvectors $\{|\nn\rangle\}_\nn$ with energy eigenvalues $\{E_\nn\}_\nn$. $H_\mathrm{int}=-q\frac{\A\cdot\pp}{m^*}$ and $H'_\mathrm{int}=q^2\frac{\A^2}{2m^*}$ are interaction Hamiltonians, and $H_\mathrm{em}=\int\!\dd^3\rr\!~\frac{\E^2+\B^2}2$ is the free electromagnetic field Hamiltonian, $\A$ being the vector potential, $\E$ and $\B$ the electric and magnetic fields. The electromagnetic field is second quantized in the Coulomb gauge \cite{Le_Bellac}. The interaction Hamiltonian $H'_\mathrm{int}$ will not be taken into account on the basis of weak intensities of light sources involved\footnote{For more details, one can refer to complement {\bf A$_\mathbf{XIII}$} of chapter {\bf XIII} in Ref. \cite{Tannoudji}.}. We now review the results of the two main approaches to the Lamb shift.

The Bethe approach to the Lamb effect is purely perturbative. The quantum second order time independent degenerate perturbation theory is applied to the Pauli-Fierz Hamiltonian $H_\mathrm{PF}$, where Hamiltonians $H_\mathrm{int}$ and $H'_\mathrm{int}$ are treated as perturbations in the weak field limit \cite{Bethe_1947}. Using renormalization arguments, the Lamb shift for an energy eigenstate $|\nn\rangle$ is found to be given by
  \begin{eqnarray}
\Delta E_\nn\lnd=\rnd\frac{2\alpha}{3\pi}\frac{q^2}{m^{*2}}\log\!\!\left(\frac{m^*}{\langle|E_\mm-E_\nn|\rangle}\right)\sum_\mm\left|\langle\mm|\pp|\nn\rangle\right|^2\!\left\{E_\mm-E_\nn\right\}\! \nt
\\
\lnd=\rnd\frac\alpha{3\pi}\frac{q^2}{m^{*2}}\log\!\!\left(\frac{m^*}{\langle|E_\mm-E_\nn|\rangle}\right)\!\langle\nn|\nabla^2V(\rr)|\nn\rangle, \label{DeltaE_lamb}
  \end{eqnarray}
where $\alpha=\frac{e^2}{4\pi}$ is the fine structure constant written in the chosen units system. $\langle|E_\mm-E_\nn|\rangle$ is the mean value of energy level difference absolute values. Historically, this predicts a Lamb shift for the hydrogen atom $2s$-level in excellent agreement with experimental values \cite{Bethe_1947, Bethe_1950}.

In the Welton approach, the Lamb shift is interpreted as a fluctuation effect on the particle position due to its interaction with the surrounding electromagnetic field. These fluctuations $\Delta\rr$ can be described as a continuous random variable, whose probability density is a three-dimensional centered isotropic Gaussian distribution of variance
  \begin{equation}
\langle(\Delta\rr)^2\rangle=\frac{2\alpha}\pi\frac{q^2}{m^{*2}}\log\!\!\left(\frac{m^*}{\kappa}\right)\!, \label{fluctuation}
  \end{equation}
where $\kappa$ is a IR cut-off, and $m^*$ is used as a natural UV cut-off consistent with non-relativistic assumption, discarding fluctuation modes of order of the particle Compton wavelength \cite{Welton_1948}. The particle then evolves in a new effective potential $\langle V(\rr+\Delta\rr)\rangle$, averaged on the fluctuation distribution, which can be written as\footnote{Even if this Taylor expansion seems to converge only for sufficiently smooth potential $V(\rr)$, it is summable in general. However, a fully quantum treatment, {\it i.e.} both particle and electromagnetic field dynamics are quantized, is needed. It has been then shown that the averaged effective potential $\langle V(\rr+\Delta\rr)\rangle$ can be obtained from the {\it bare} potential $V(\rr)$ by applying the well-defined differential operator \cite{Welton_1948}
  $$
\langle V(\rr+\Delta\rr)\rangle=\e^{\frac{\langle(\Delta\rr)^2\rangle}6\nabla^2}V(\rr).
  $$}
  $$
\langle V(\rr+\Delta\rr)\rangle=\left\{1+\frac{\langle(\Delta\rr)^2\rangle}6\nabla^2+\dots\right\}\!V(\rr)=V(\rr)+\Delta V(\rr).
  $$
where ellipsis dots ``\dots'' denote terms of order higher than first order in $\alpha$. The corrective term $\Delta V(\rr)$ of the first order in $\alpha$ is precisely the term giving rise to the Lamb shift
  $$
\Delta E_\nn=\int\!\dd^3\rr\left|\langle\rr|\nn\rangle\right|^2\Delta V(\rr)=\frac\alpha{3\pi}\frac{q^2}{m^{*2}}\log\!\!\left(\frac{m^*}\kappa\right)\!\langle\nn|\nabla^2V(\rr)|\nn\rangle.
  $$
Comparison of this result with Eq. (\ref{DeltaE_lamb}) shows that the IR cut-off $\kappa$ can be identified to $\langle|E_\mm-E_\nn|\rangle$ in the Bethe result. This also means that $\kappa=\langle|E_\mm-E_\nn|\rangle$ should not depend on the quantum numbers $\nn$ \cite{Welton_1948}.

\section{The physics of a spherical semiconductor quantum dot} \label{sec_3}
First, a QD may be considered as spherical if it possesses an aspect ratio, defined as the ratio of its longest over its shortest axe, smaller than 1.1. For example, according to the growth scheme used to synthesize $CdSe$ nanocrystals, one can produce spherical QDs of radius in the range of a few tens \AA~\cite{Millo_2004} or even less \cite{Reiss_2002}, as well as rod-shaped QDs, so-called nanorods, having an aspect ratio up to more than 10 \cite{Peng_2002}. The general issue of the shape of semiconducting QDs, and how QDs of different shapes should be experimentaly obtained by appropriate growth technique, is extensively treated in Ref. \cite{Kudera_2008}.

Next, it is admitted that, under reasonable physical assumptions, semiconductor QDs may be described by EMA models in a first approximation. The most important one is the parabolicity of the semiconductor band structure \cite{Brus_1984}. This assumption may be efficiently amended in several ways. A review presenting the usual possibilities to do so is given in Ref. \cite{Yoffe_2002}. Here, we use the standard EMA model presented in Ref. \cite{Billaud_2009a}, as a first attempt to apprehend the Lamb effect in semiconductor QDs. Electrons and holes are assumed to be non-relativistic spinless particles of effective masses $m_{\e,\h}^*$, confined in an infinite spherical potential well
  $$
V^\infty(\rr)=\left\{\begin{array}{cccl}
0 & \textrm{if} & 0\leq r<R, & \textrm{region I~};
\\
\infty & \textrm{if} & r\geq R, & \textrm{region II}.
\end{array}
\right.
  $$
The electron-hole Coulomb interaction is to be taken into account subsequently by the Ritz variational principle. But, they are also consequently isolated from the insulating surrounding of the QD. The electromagnetic field amplitude should not then exceed some threshold, so that charge carriers would not leak out by tunnel effect. This working assumption is known as the weak field limit. Such model leads to an overestimation of the electron-hole pair ground state energy for small QDs, which can be corrected by restoring a confining finite potential step of experimentally acceptable height \cite{Thoai_1990}. Other QD models with parabolic confinement \cite{Keller_1995, Lozovik_2003} or parabolic potential superimposed to an infinite potential well \cite{Sundqvist_2002} exist. But, the concept of QD size is no longer well-defined since their eigenfunctions are delocalized.
\subsection{Interactive electron-hole pair EMA model} \label{subsec_3_1}
The electron-hole Coulomb interaction $V_\C(\rr_{\e\h})$ is taken into account in the total Hamiltonian
  $$
H_0=H_\e+H_\h+V_\C(\rr_{\e\h}),
  $$
where $\kappa=4\pi\varepsilon$, $\varepsilon$ denotes the semiconductor dielectric constant, $r_{\e\h}$ the electron-hole relative distance, and $H_{\e,\h}=-\frac{\nabla^2_{\!\e,\h}}{2m_{\e,\h}^*}+V^\infty(\rr_{\e,\h})$ the electron and hole confinement Hamiltonian. Without loss of generality, the semiconductor energy band gap $E_\g$ may be set equal to be zero for convenience. Electron and hole, as decoupled particles, have wave functions of the form
  \begin{eqnarray*}
\psi^\infty_{lnm}(\rr_{\e,\h})\lnd=\rnd R_{ln}^{\I\infty}(r_{\e,\h})Y^m_l(\theta_{\e,\h},\varphi_{\e,\h})
\\
\lnd=\rnd\sqrt{\frac2{R^3}}\frac{\chi_{[0,R[}(r_{\e,\h})}{\j_{l+1}(k^\infty_{ln})}\j_l\!\!\left(\frac{k^\infty_{ln}}Rr_{\e,\h}\!\right)\!\Y^m_l(\theta_{\e,\h},\varphi_{\e,\h}).
  \end{eqnarray*}
where $l\in\mathbb N$, $n\in\mathbb N\smallsetminus\{0\}$ and $m\in\ldc-l,l\rdc$, $\Y^m_l(\theta,\varphi)$ are spherical harmonics and $\j_l(x)$ spherical Bessel functions of the first kind, $\chi_{\mathbb A}(r)=\left\{
  \begin{array}{cl}
1 & \textrm{if}~~r\in\mathbb A
\\
0 & \textrm{otherwise}
  \end{array}
\right.$ the radial characteristic function of the set $\mathbb A$. Finally, $\left\{k^\infty_{ln}\right\}_{ln}$ is the wave numbers set, defined as the $n^\mathrm{th}$ non-zero root of $\j_l(x)$, from the continuity conditions at $r=R$ \cite{Efros_1982}. The respective energy eigenvalues for electron and hole are expressed in terms of $\{k^\infty_{ln}\}_{ln}$ as
  $$
E^{\e,\h\infty}_{ln}=\frac{(k^\infty_{ln})^2}{2m^*_{\e,\h}R^2}.
  $$

The interplay between quantum confinement energy, scaling as $\propto R^{-2}$, and Coulomb interaction, scaling as $\propto R^{-1}$, can be described by the ratio $\frac R{a^*}$ of the QD radius $R$ over the Bohr radius of the bulk Mott-Wannier exciton $a^*=\frac{\kappa}{e^2\mu}$, $\mu$ being the reduced mass of the exciton. As an exact analytic solution is beyond reach, two regimes are discussed in Ref. \cite{Kayanuma_1988}.
\subsection{Strong confinement regime} \label{subsec_3_2}
In this regime, valid for $R\lesssim 2a^*$, the electron-hole relative motion is so affected by the infinite potential well, that {\it exciton} states should be considered as uncorrelated electronic and hole states. The Coulomb potential is considered as a perturbation with respect to the infinite confining potential well. A variational approach is used to obtain the electron-hole pair ground state, with the following trial function
  \begin{equation}
\phi^\infty(\rr_\e,\rr_\h)=\psi^\infty_{010}(\rr_\e)\psi^\infty_{010}(\rr_\h)\phi_\mathrm{rel}(\rr_{\e\h}), \label{phi}
  \end{equation}
which is a product of non-interacting charge carriers ground state wave functions with an interaction wave function of the form $\phi_\mathrm{rel}(\rr_{\e\h})=\ee^{-\frac\sigma2r_{\e\h}}$, $\sigma$ being a variational parameter. The electron-hole pair energy minimization selects the value $\sigma_0=\frac{4B'}{a^*}$, and yields the ground state energy\footnote{All constants appearing in the text and formulas are listed in appendix {\bf\ref{appendix_A}}.}
  $$
E^\mathrm{strong}_{\e\h}=E^\infty_{\e\h}-A\frac{e^2}{\kappa R}-4B'^2E^*,
  $$
where $\ds E^\infty_{\e\h}=E^{\e\infty}_{01}+E^{\h\infty}_{01}$ is the electron-hole pair ground state confinement energy, and $E^*=\frac1{2\mu a^{*2}}$ the binding exciton Rydberg energy \cite{Kayanuma_1988}.
\subsection{Weak confinement regime} \label{subsec_3_3}
In this regime, valid for $R\gtrsim 4a^*$, the exciton retains its character of a quasi-particle of total mass $M=m^*_\e+m^*_\h$. Its center-of-mass motion is confined, and should be quantized. The Coulomb interaction remains a perturbation to the infinite confining potential well, and the variational function $\phi^\infty(\rr_\e,\rr_\h)$ should be kept. However, the QD size allows for a partial restoration of the long range Coulomb potential between the charged carriers, such that the Coulomb energy and the kinetic energy in electron-hole relative coordinates are of the same order of magnitude. Then, the leading contribution to the ground state energy of the exciton should be $-E^*$, which may be viewed as the ground state energy of a hydrogen-like atom of mass $\mu$. The total translational motion of the exciton, should be restored and contribute to the exciton total energy by a term $\frac{\pi^2}{2MR^2}$, {\it i.e.} the ground state energy of a free particle trapped in a region of size $R$.

The exciton center-of-mass cannot reach the infinite potential well surface unless the electron-hole relative motion undergoes a strong deformation. The exciton should be preferably treated as a rigid sphere of radius $\eta(\lambda)a^*$, where $\eta(\lambda)$ is a phenomenologically determined function of $\lambda=\frac{m_\h^*}{m_\e^*}$ \cite{Kayanuma_1988}. On this basis, to improve the description of the excitonic ground state, we let the center of mass motion in the variational procedure be represented by a plane wave $\phi_\G(\rr_\G)=\ee^{i\frac\pi R\textrm{\scriptsize$\sigmag$}_\G\cdot\rr_\G}$, where $\rr_\G$ is the center-of-mass coordinates and $\sigmag_\G$ is vector quantum number of unit norm $|\sigmag_\G|^2=1$. The trial function $\phi^\infty(\rr_\e,\rr_\h)$ takes now the form
  \begin{equation}
\psi^\infty(\rr_\e,\rr_\h)=\phi^\infty(\rr_\e,\rr_\h)\phi_\G(\rr_\G). \label{psi}
  \end{equation}
This leaves the exciton probability density unchanged as well as the Coulomb potential matrix element, whereas the confinement Hamiltonian $H_\e+H_\h$ mean value gets the appropriate further contribution $\frac{\pi^2}{2MR^2}$. The variational calculation yields $\sigma_0\approx\ds2a^{*-1}$, and a ground state variational energy of
  \begin{equation}
E^\mathrm{weak}_{\e\h}=-E^*+\frac{\pi^2}{6\mu R^2}+\frac{\pi^2}{2M(R-\eta(\lambda)a^*)^2}, \label{E^weak_eh}
  \end{equation}
where terms of fourth and higher order terms in $\frac{a^*}R$ are neglected. An analytical expression for $\eta(\lambda)$ can now be extracted from Eq. (\ref{E^weak_eh}), as shown in Ref. \cite{Billaud_2009a}.
\subsection{Pseudo-potential-like method}
The exciton ground state energy $E^\mathrm{weak}_{\e\h}$ of \cite{Kayanuma_1988}, differs from Eq. (\ref{E^weak_eh}) by the term $\frac{\pi^2}{6\mu R^2}$, interpreted as a kinetic energy term in the relative coordinates. As the virial theorem in this set of coordinates should be satisfied, this energy should be already contained in the Rydberg energy term $-E^*$, and therefore should be removed. An elegant way to do this consists in introducing a pseudo-potential
  $$
W(\rr_{\e\h})=W\frac{r_{\e\h}^2}{R^2}\ee^{-2\frac{r_{\e\h}}{a^*}}=-\frac{32\pi^2}9E^*\frac{r_{\e\h}^2}{R^2}\ee^{-2\frac{r_{\e\h}}{a^*}}.
  $$
Inspection shows that it makes contributions to the second order of the exciton total energy but not to the third one, while leaving the zeroth and first order terms. Higher order contributions are interpreted as higher order corrections to the kinetic energy of the exciton $\frac{\pi^2}{2MR^2}$. While the amplitude $W$ is to be fixed to get the correct kinetic energy $\frac{\langle\psi|W(\rr_{\e\h})|\psi\rangle}{\langle\psi|\psi\rangle}=-\frac{\pi^2}{6\mu R^2}\!\left\{1+O\!\left(\frac{a^{*2}}{R^2}\right)\!\right\}$, the pseudo-potential form is not arbitrary. It is attractive at distances $\approx a^*$ to promote excitonic state with typical size around its Bohr radius, repulsive at short distances to penalize excitonic state with small size, and exponentially small for large distances in order not to perturb the long range Coulomb potential \cite{Billaud_2009a}.

Adding the pseudo-potential $W(\rr_{\e\h})$ to the Hamiltonian $H_0$ implies a significant decrease of the expected value of the exciton energy in the strong confinement regime $\frac{\langle\phi|W(\rr_{\e\h})|\phi\rangle}{\langle\phi|\phi\rangle}=-\frac{64\pi^2}9CE^*\!\left\{1+O\!\left(\frac R{a^*}\right)\!\right\}$, which is now only valid for $2R\lesssim a^*$. The excitonic energy computed in presence of the pseudo-potential gets a better fit to experimental results in this validity domain, than those calculated without this tool. Nevertheless, the divergence for very small QD size still persists as a relic of the infinite potential well assumption.
\section{Lamb shift of a particle confined in a spherical potential well} \label{sec_4}
This section deals with the Lamb shift for a particle of mass $m^*$ and of charge $\pm e$, {\it i.e.} $q=\pm1$, confined in a {\it theoretical} spherical QD with a spherical infinite potential well $V^\infty(\rr)$\footnote{The notation $X^\infty=\ds\lim_{V\infty}X$ is adopted, for all physical quantities. This is the reason why we will choose to note the infinite potential well $V^\infty(\rr)$, and the finite potential step $V(\rr)$, in sections {\bf\ref{sec_4}} and {\bf\ref{sec_5}}.}. The difficulty then resides in the determination of the Poisson equation satisfied by this singular potential. To overcome this issue, we will replace it by a potential step of finite constant height $V\geq0$
  $$
V(\rr)=V\chi_{]R,\infty[}(r)=\left\{\begin{array}{cccl}
0 & \textrm{if} & 0\leq r<R, & \textrm{region I~};
\\
V & \textrm{if} & r\geq R, & \textrm{region II}.
\end{array}
\right.
  $$
In the limit $V\rightarrow\infty$, the Lamb shift undergone by an energy level of the particle in the infinite potential well is supposed to be the finite part of the expansion in powers of $V$ of the Lamb shift undergone by the level with same quantum numbers of the particle in the finite potential step, given by Eq. (\ref{DeltaE_lamb})
  $$
\Delta E_{lnm}=\frac\alpha{3\pi m^{*2}}\log\!\!\left(\frac{m^*}{\kappa}\right)\!\langle\psi_{lnm}|\nabla^2V(\rr)|\psi_{lnm}\rangle,
  $$
where, for fixed quantum numbers $l\in\mathbb N$, $n\in\mathbb N\smallsetminus\{0\}$ and $m\in\ldc-l,l\rdc$, the wave function $\psi_{lnm}(\rr)$ is the eigen function of the Hamiltonian $H_0$ of the particle confined by the finite potential step $V(\rr)$ instead of the infinite potential well $V^\infty(\rr)$\footnote{For more details, one can see appendix {\bf\ref{appendix_B}}. In particular, $\psi_{lnm}(\rr)$ is a linear combination of the spherical Bessel function $\j_l\!\!\left(\frac{k_{ln}}Rr\right)$ and of the spherical Hankel function $\h^{(1)}_l\!\!\left(i\frac{K_{ln}}Rr\right)$, $k_{ln}$ and $K_{ln}$ being the wave numbers in the regions I and II respectively.}.
\subsection{Lamb shift} \label{subsec_4_1}
In this new formalism, it becomes possible to find the Poisson equation satisfied by the potential step $V(\rr)$, at least in the distributions sense. The Lamb shift of any quantum state of the confined particle is therefore obtained as
  \begin{equation}
\Delta E^\infty_{ln}=-\frac8{3\pi}\frac\alpha{m^{*2}}\frac{E_{ln}^\infty}{R^2}\log\!\!\left(\frac{m^*}{\kappa}\right)\!\leq0. \label{DeltaE^infty_lamb}
  \end{equation}
More details on this calculation is given in appendix {\it\ref{appendix_B_1}}. As expected, because of the spherical symmetry of the potentials $V(\rr)$ and $V^\infty(\rr)$, the Lamb shift is independent of the azimuthal quantum number $m$. Furthermore, it is {\it negative}, since, as we shall see later, the IR cut-off satisfies the constraint $\kappa\leq m^*$. Hence, the Lamb effect brings down the energy levels of the particle, instead of raising them up, as in real atoms. This is a remarkable characteristic of the Lamb shift which, to the best of our knowledge, is found for the first time.

In appendix {\it\ref{appendix_B_2}}, we present an alternative calculation of this Lamb shift. It is based on a regularized but non-resumed version of Eq. (\ref{DeltaE_lamb}), and leads to a Lamb shift, identical to Eq. (\ref{DeltaE^infty_lamb}). Despite an identical regularization method, no divergent terms appear in the limit $V\rightarrow\infty$, which is most satisfactory from the conceptual point of view. This corroborates the fact that such a term is not physical, when a trapped particle is confined by the infinite potential well $V^\infty(\rr)$. As we will see later, this property appears to be fundamental. Finally, this second calculation validates the regularization of the confining potential by an intermediate finite potential step.
\subsection{Infrared cut-off $\kappa$} \label{subsec_4_2}
The evaluation of the IR cut-off $\kappa=\left\langle|E^\infty_{ln}-E^\infty_{ij}|\right\rangle$ is also problematic. In the Bethe approach, the IR cut-off explicitly depends on the considered quantum state, which is not the case in the Welton approach. In real atoms, it has been shown that the IR cut-off introduced by Bethe in his original article is actually of about the same order of magnitude independently of the energy levels \cite{Bethe_1950}. Therefore, to introduce a appropriate IR cut-off $\kappa$, independent from quantum numbers of the considered energy level, as required by the Welton approach, we suggest to define it by the formal average of the quantum state dependent IR cut-offs of the Bethe approach over all possible quantum numbers \cite{Bethe_1950}
  \begin{equation}
\kappa=\left\langle|E^\infty_{ln}-E^\infty_{ij}|\right\rangle=\frac{\sum_{ijklnm}|E^\infty_{ln}-E^\infty_{ij}|}{\sum_{ijklnm}1}=\frac1{2m^*R^2}\frac{\sum_{ijln}(2i+1)(2l+1)|(k^\infty_{ln})^2-(k^\infty_{ij})^2|}{\sum_{ijln}(2i+1)(2l+1)}. \label{kappa}
  \end{equation}
As quantum numbers are not limited from bellow, and are infinitely many, the sum of Eq. (\ref{kappa}) is infinite. To confer a rigorous meaning to Eq. (\ref{kappa}), a regularization method should be prescribed. Actually, the number of terms in the sum is finite, since the trapped particle, being excited through its interaction with the electromagnetic field, can have access only to a finite number of states. As the highest accessible level is limited by the finite field energy $E_\mathrm{lim}=\frac{\kappa_\mathrm{lim}^2}{2m^*R^2}$, there is a way to define a {\it UV cut-off} $\kappa_\mathrm{lim}$ for authorized wave numbers. In practice, we make use of the so-called Feynman regularization method, presented in appendix {\bf\ref{appendix_C}}, which leads to the choice
  $$
\kappa=\frac{7\pi^2}{12m^*R^2}.
  $$
Thus, $\kappa$ is of the order of magnitude of the ground state energy of the confined particle, contrary to the case of real atoms \cite{Bethe_1950}. Because of the state degeneracy, such behavior is {\it a posteriori} expected. Terms, whose relative importance is the largest in Eq. (\ref{kappa}), are terms of the order of magnitude $\approx|(k^\infty_{ln+1})^2-(k^\infty_{ln})^2|$. The asymptotical behavior of the Bessel functions roots yields $|(k^\infty_{ln+1})^2-(k^\infty_{ln})^2|\approx\pi^2(2n+1)$, for sufficiently large $n\in\mathbb N\smallsetminus\{0\}$. By regularizing sums over $n$, since sums over $l$ do not play any role, we deduce that $\kappa\approx\frac{\pi^2}{m^*R^2}$.

The non-relativistic condition in the IR and UV cut-offs $\kappa(R)\leq m^*$ imposes that there exists a lower potential radius bound $R^*_\mathrm{min}=\frac\pi2\sqrt{\frac73}\lambdabarre^*\approx2,399\lambdabarre^*$, of the same order of magnitude of the particle reduced Compton wave length $\lambdabarre^*=m^{*-1}$. Thus, in a {\it theoretical} QD of radius $R$ smaller than $R^*_\mathrm{min}$, the confined particle should acquire at least a confinement energy $E^\infty_{01}$ of the order of magnitude of its rest mass energy $m^*\propto\frac{\pi^2}{2m^*R^{*2}_\mathrm{min}}\leq\frac{\pi^2}{2m^*R^2}=E^\infty_{01}$, which will explicitly contravene to the non-relativistic assumption. Finally, there exists another particular potential radius $R^*_\mathrm{max}=\sqrt\e R^*_\mathrm{min}\approx3,956\lambdabarre^*$, for which the Lamb shift of any energy level represents a maximal fraction of it, {\it i.e.} $\frac\dd{\dd R}\frac{\Delta E^\infty_{ln}}{E^\infty_{ln}}\Big|_{R=R^*_\mathrm{max}}=0$.
\section{Gauge invariance and Lamb shift} \label{sec_5}
As an observable effect, the Lamb shift is indeed gauge invariant. Welton argument is clearly gauge invariant, since the electromagnetic field only intervenes through its energy density, a gauge invariant quantity. Bethe argument depends however on the gauge, since the interaction Hamiltonian $H_\mathrm{int}\propto\A\cdot\pp$ in the Pauli-Fierz Hamiltonian $H_\mathrm{PF}$, used as perturbation, is written in the Coulomb gauge. Therefore, it is useful to check gauge invariance in calculations of the Lamb shift of the energy levels for a {\it theoretical} QD. To this end, we study the Lamb effect in the gauge imposed by the electric dipole approximation.

In this approximation, the electric field spatial variation is negligible over typical particle distances --- {\it i.e.} $\E=\E(t,\zerog)$ and the related scalar potential is $A^0(t,\rr)=\ds-\rr\cdot\E(t,\zerog)$ ---, and that there is no magnetic field $\B=\zerog$, hence no vector potential $\A=\zerog$. The system Hamiltonian is then of the form
  $$
H'_\mathrm{PF}=H_0+H_\mathrm{em}+eH''_\mathrm{int},
  $$
where $H''_\mathrm{int}=-q\rr\cdot\E(t,\zerog)$ is the new interaction Hamiltonian of the particle with the electromagnetic field. In particular, it is obvious to observe that $H_\mathrm{int}$ and $H''_\mathrm{int}$ contribute in the same way to the classical action of the particle-electromagnetic field system, especially when quadratic terms in $\A$ are neglected\footnote{By a simple integration by parts, one can show that $\int\!\dd t\!~H_\mathrm{int}=\!\int\!\dd t\!~H''_\mathrm{int}$.}. Thus, quantization in the gauge fixed by the electric dipole approximation makes sense.

We now go over Bethe argument with the dipole approximation. The second order correction is
   \begin{eqnarray}
\Delta E_\nn\lnd=\rnd-\frac{2\alpha}{3\pi}q^2\sum_\mm\left|\langle\mm|\rr|\nn\rangle\right|^2\!\!\left\{\int\!\dd k\!\left[k^2-(E_\mm-E_\nn)k+(E_\mm-E_\nn)^2\right]-\int\!\dd k\frac{(E_\mm-E_\nn)^3}{E_\mm-E_\nn+k}\right\} \nt
\\
\lnd=\rnd-\frac{2\alpha}{3\pi}q^2\sum_\mm\left|\langle\mm|\rr|\nn\rangle\right|^2\!\left\{\frac{m^{*3}}3-(E_\mm-E_\nn)\frac{m^{*2}}2+(E_\mm-E_\nn)^2m^*-(E_\mm-E_\nn)^3\log\!\!\left(\frac{m^*}\kappa\right)\!\right\}\!. \label{DeltaE_lamb'}
  \end{eqnarray}
The two integrals in this expression have respectively a cubic and a logarithmic UV divergence. Thus, the second integral is the one which gives rise to the Lamb shift in the dipole approximation. It is natural to introduce $m^*$ as UV cut-off and $\kappa$ as IR cut-off to regularize these integrals. Renormalization arguments of the original Bethe article are used to get the corrective term
  $$
\Delta E_\nn=-\frac{2\alpha}{9\pi}q^2m^{*3}\langle\nn|\rr^2|\nn\rangle+\frac{\alpha}{2\pi}q^2m^*+\frac\alpha{3\pi}\frac{q^2}{m^{*2}}\log\!\!\left(\frac{m^*}\kappa\right)\!\!\langle\nn|\nabla^2V(\rr)|\nn\rangle.
  $$
The third term in $\Delta E_\nn$ is recognized as the Lamb shift for a state $|\nn\rangle$, as expected from the Welton approach or the Bethe approach in the Coulomb gauge. The second term is simply a numerical constant, which can be omitted using an additive renormalization argument. The first term is explicitly gauge dependent, since $\langle\nn|\rr^2|\nn\rangle\propto\langle\nn|H''^2_\mathrm{int}|\nn\rangle=\sum_\mm|\langle\mm|H''_\mathrm{int}|\nn\rangle|^2$. So, in a general way, the Lamb shift in the gauge fixed by the electric dipole approximation is identical to the one computed in the Coulomb gauge.

This should be explicitly verified in the case of a particle confined in the infinite potential well $V^\infty(\rr)$. As described in more details in appendix {\it\ref{appendix_B_3}}, we show that the regularized version by the finite potential step $V(\rr)$ of Eq. (\ref{DeltaE_lamb'}), but non-resummed by the use of the closure relation, leads to the same Lamb shift than the one predicted by Eq. (\ref{DeltaE^infty_lamb}) for any quantum state of the confined particle. As mentioned in subsection {\it\ref{subsec_4_1}}, this alternative calculation of its Lamb shift also eliminates divergent terms in the limit $V\rightarrow\infty$. This reinforces once again the argument according to which the term scaling as $\propto\sqrt V$, which appears in the calculation method presented in subsection {\it\ref{subsec_4_1}}, is an artefact of the regularization method, and should be dropped.
\section{Comments on the Lamb shift of a particle confined in a spherical potential well} \label{sec_6}
The observed negative sign of the Lamb shift for a particle confined in a spherical potential found in subsection {\it\ref{subsec_4_1}}, as opposed to positive Lamb shifts known for hydrogen-like atoms, calls for a comment. We first present a heuristic argument, and then back it up with a mathematical approach.
\subsection{A heuristic view}
In Quantum Electrodynamics (QED), the quantized electromagnetic field is represented by an assembly of harmonic oscillators, which are known to have a non-zero ground state energy, often called the zero-point energy. Even in absence of any external radiation field, non-zero electromagnetic fields are present as arising from the zero-point energy of all the associated oscillators. The effect of these {\it vacuum fluctuations} on a charged particle is to smear out its position causing a spread of charge and mass in space, characterized by a typical radius of the order of magnitude of the deviation $\sqrt{\langle(\Delta \rr)^2\rangle}$ \cite{Eides_2001}.

In hydrogen-like atoms, the spreading effect generates a correction to the Coulomb potential. The electron charge is distributed over a larger volume of space and less of its charge experience the nuclear Coulomb potential, resulting in a reduction of its binding energy to the nucleus. Electron $s$-states, being closer to the nucleus than other states, would experience a larger reduction of the binding energy. Hence, the major effect of the vacuum fluctuations is to shift $s$-states upward in energy relative to $p$-states. This energy difference is predicted by QED to be of about $+1086$ MHz for the hydrogen atom 2$s$ state, which is the leading contribution to the observed Lamb shift \cite{Eides_2001}.

The situation is different in a {\it theoretical} QD, since no force acts on the charged particle inside the QD the apart infinite wall reflection at its boundary. Thus, the observed effect is not due to an electric charge spreading but to a mass spreading. In this situation, the total energy which was concentrated in the kinetic energy of the point-like particle is now transferred to the energy of a spatial mass distribution, which splits into center of mass motion and relative motion. On the basis of total energy conservation, one should expect then a {\it reduction} of the center mass motion energy, an effect which is opposite to the one observed in atoms.

Among the secondary contributions to the Lamb shift, it is the vacuum polarization which is of largest order. Virtual pairs of electrons and positrons are continuously created and annihilated in vacuum.

Inside an hydrogen-like atom, virtual electrons are attracted to the nucleus, while positrons are repelled. The result is that the nucleus charge is screened by the virtual electrons. But, a bound electron inside the cloud of virtual electrons will experience a stronger attraction to the nucleus than predicted by the standard Coulomb potential, and hence will be more tightly bound to the nucleus. Since $s$-state electrons are closer to the nucleus than other electrons, they will experience a larger increase in binding energy, which can be evaluated by QED to be of about $-27$ MHz for the hydrogen atom 2$s$ state \cite{Eides_2001}.

But, inside a QD, the charged particle appears to move in a polarized medium in the mean. Thus, its motion will be kinematically enhanced, and the corresponding energy level raised a bit. In short, the Lamb shift in QDs is of very different nature as compared to usual Lamb shift in atoms.
\subsection{A mathematical approach}
In the calculation of the Lamb shift, the main object is the series appearing in Eq. (\ref{DeltaE_lamb}), which we write for an hydrogen-like atom as $\sum_{ijk}\left|\langle\Psi_{ijk}|\pp|\Psi_{nlm}\rangle\right|^2\!\left\{E_i-E_n\right\}$, where  $\Psi_{nlm}(\rr)$ and  $E_n$ are the eigenfunction and the corresponding energy for a state labeled by the quantum numbers $n\in\mathbb N\smallsetminus\{0\}$, $l\in\ldc0,n-1\rdc$ and $m\in\ldc-l,l\rdc$\footnote{For more details, one can refer to chapter {\bf VII} in Ref. \cite{Tannoudji}.}. This series is indeed convergent because of the closure relation and because $E_n\propto\frac1{n^2}$, implying that $E_n\xrightarrow[n\infty]{}0$. Its sum can be shown to be $\frac12\langle\Psi_{nlm}|\nabla^2V_\C(\rr)|\Psi_{nlm}\rangle\propto|\Psi_{nlm}(\zerog)|^2\geq0$. The same reasoning holds for the series $\sum_{ijk}\left|\langle\Psi_{ijk}|\rr|\Psi_{nlm}\rangle\right|^2(E_i-E_n)^3=\frac1{2m_\e^2}\langle\Psi_{nlm}|\nabla^2V_\C(\rr)|\Psi_{nlm}\rangle$ appearing in Eq. (\ref{DeltaE_lamb'}). This shows why the atomic Lamb shift is positive.

For a QD, the two relevant series $\sum_{ijk}\left|\langle\psi_{ijk}|\pp|\psi_{lnm}\rangle\right|^2\!\left\{E_{ij}-E_{ln}\right\}$ and $\sum_{ijk}\left|\langle\psi_{ijk}|\rr|\psi_{lnm}\rangle\right|^2(E_{ij}-E_{ln})^3$ have a completely different behavior. They are indeed not well defined, even if the regularization finite potential step $V(\rr)$ is introduced, since the series $\sum_{ijk}\left|\langle\psi_{ijk}|\pp|\psi_{nlm}\rangle\right|^2E^p_{ij}~(p=1,2,3)$ are divergent, because $E_{ln}\propto k^2_{ln}\xrightarrow[l,n\infty]{}\infty$. Yet, we have to extract a finite value from these formal series and there is no way {\it a priori} to predict the sign of the finite term\footnote{This is a well known problem in divergent series regularization. For example, the Feynman regularization method introduced in appendix {\it\ref{appendix_B_3}} allows to show that $\sum_{p\geq1}p^2=-\frac1{12}\leq0$. This negative extracted value is coherent with the value of the analytical extension of the Riemann function $\zeta(z)=\sum_{p\geq1}\frac1{p^z}$ to $\mathbb C\smallsetminus\{1\}$ \cite{Abramowitz}.}. To confer a proper meaning to these series, divergences should be removed by using the definition of the eigen-energies, {\it i.e.} $H_0|\psi_{lnm}\rangle=E_{ln}|\psi_{lnm}\rangle$, and the closure relation, which stays valid because the eigenfunctions $\psi_{lnm}(\rr)$ are orthonormalized. This yields $\sum_{ijk}\left|\langle\psi_{ijk}|\pp|\psi_{lnm}\rangle\right|^2\{E_{ij}-E_{ln}\}=-\langle\psi_{lnm}|\nabla\cdot H_0\nabla-E_{ln}\nabla^2|\psi_{lnm}\rangle$ and $\sum_{ijk}\left|\langle\psi_{ijk}|\rr|\psi_{lnm}\rangle\right|^2(E_{ij}-E_{ln})^3=\langle\psi_{lnm}|[H_0,\rr]^\dag\!\cdot\!\left[H_0,[H_0,\rr]\right]\!|\psi_{lnm}\rangle$, as shown in appendixes {\it\ref{appendix_B_2}} and {\it\ref{appendix_B_3}}.

The validity of this approach is confirmed since their expansion in powers of the potential step height $V$ does not contain divergent terms in the limit $V\rightarrow\infty$. It also allows to understand why the expansion in powers of $V$ of the matrix element $\langle\psi_{lnm}|\nabla^2V(\rr)|\psi_{lnm}\rangle$ contains a divergent term behaving as $\propto\sqrt V$. Intuitively, the usual commutator relations are no longer valid, because the confined particle can have only access to a finite spatial region with a probability arbitrarily close to one, in the limit $V\rightarrow\infty$. Thus, the effect of the surface at the boundary $r=R$ has not been taken into account satisfactorily. If the commutation relations are not used, this problem vanishes.

The sign of the Lamb shift of the confined particle in the {\it theoretical} QD is then easily predictable. Sections {\bf\ref{sec_4}} and {\bf\ref{sec_5}} suggests that it is the behavior of the radial wave function on the sphere $r=R$, which is responsible for the Lamb shift. It can be seen, from expressions of appendix {\bf\ref{appendix_B}} that, since the contribution due to the Dirac distribution $\delta(r-R)$ vanishes in the limit $V\rightarrow\infty$ because of the continuity condition $R^{\I\infty}_{ln}(R)=0$, the contribution due to the Dirac distribution derivative $\delta'(r-R)$ is then the crucial one. Since $R^{\I\infty\prime}_{ln}(R)\neq0$ and $R^{\I\infty\prime\prime}_{ln}(R)\neq0$, using standard properties of the spherical Bessel functions in a neighborhood of one of its roots, $R^{\I\infty\prime}_{ln}(R)$ and $R^{\I\infty\prime\prime}_{ln}(R)$ are of opposite sign. From this fact, it follows that the QD Lamb shift is negative.
\section{Lamb shift in spherical semiconductor quantum dots} \label{sec_7}
In a spherical semiconductor QD, the interaction of the electron-hole pair with the quantized electromagnetic field generates a Lamb shift in its energy levels. In the Welton approach, electron and hole positions are fluctuating. Consequently, the Lamb shift of the electron-hole pair ground state consists of the sum of the contributions of Lamb shift undergone by the electron and by the hole, when the pair is in its ground state. It will be evaluated for each confinement regime of section {\bf\ref{sec_3}}. The electron and hole respective position fluctuation variance
  $$
\langle(\Delta\rr_{\e,\h})^2\rangle=\frac{2\alpha}{\pi\varepsilon}\frac1{m_{\e,\h}^{*2}}\log\!\! \left(\frac{m^*_{\e,\h}}{\kappa^*_{\e,\h}}\right)\!,
  $$
are given by expressions analogous to Eq. (\ref{fluctuation}), where $\kappa^*_{\e,\h}$ are electron and hole IR cut-offs. Then, in a semiconductor for which the fluctuations are large enough, {\it i.e.} for which the effective masses $m^*_{\e,\h}$ are significantly smaller than the electron bare mass $m_\e$, there exists a possibility to detect experimentally the Lamb shift for a range of reasonable QD sizes.

In this section, we perform Taylor expansions of the Lamb shift in both confinement regimes to the second order terms to account for the pseudo-potential $W(\rr_{\e\h})$. As the contributions to the electron-hole pair confinement energy come only from second order terms, diagonal matrix elements of the action of the Laplacian on potentials are to be evaluated to this order. Therefore, the Lamb shift undergone by the electron or by the hole has three contributions. The first one is a boundary effect due to the confinement potential well $V^\infty(\rr_{\e,\h})$, which is of the same nature as the one experienced by a confined particle in sections {\bf\ref{sec_4}} or {\bf\ref{sec_5}}. The second one is due to the Coulomb interaction $V_\C(\rr_{\e\h})$ between the electron and the hole, analogous to the Lamb shift in real atoms. And, finally, the third comes from the pseudo-potential $W(\rr_{\e\h})$.
\subsection{General considerations}
Let the non-normalized wave functions of the electron-hole pair ground state, confined by the step potential $V(\rr_{\e,\h})$ in both confinement regimes of Eqs. (\ref{phi}) and (\ref{psi}) be
  $$
\phi(\rr_\e,\rr_\h)=\psi^\e_{010}(\rr_\e)\psi^\h_{010}(\rr_\h)\phi_{\e\h}(\rr_{\e\h}) \mathrm{~~~~and~~~~}\psi(\rr_\e,\rr_\h)=\psi(\rr_\e,\rr_\h)\phi_\G(\rr_\G).
  $$
Here, electron and hole ground state wave functions confined individually $\psi^{\e,\h}_{010}(\rr_{\e,\h})$ depend on the quantum numbers $k^{\e,\h}_{01}$ and $K^{\e,\h}_{01}$, subjected to $k^{\e,\h2}_{01}+K^{\e,\h2}_{01}=2m^*_{\e,\h}R^2V$. The function $\phi_\G(\rr_\G)$ is a pure phase factor and can be dropped. Since the variational parameter $\sigma$ appears in $\phi^\infty(\rr_\e,\rr_\h)$, it should be replaced by its variational value $\sigma_0$, proper to each confinement regime. Then, the electron or the hole Lamb shift is
  $$
\Delta E^\infty_{\e,\h}=\frac{\alpha}{3\pi\varepsilon}\frac1{m_{\e,\h}^{*2}}\log\!\!\left(\frac{m^*_{\e,\h}}{\kappa^*_{\e,\h}}\right)\!\!\left\{\frac{\langle\phi^\infty|\nabla^2V^\infty(\rr_{\e,\h})|\phi^\infty\rangle}{\langle\phi^\infty|\phi^\infty\rangle}+\frac{\langle\phi^\infty|\nabla^2V_\C(\rr_{\e\h})|\phi^\infty\rangle}{\langle\phi^\infty|\phi^\infty\rangle}+\frac{\langle\phi^\infty|\nabla^2W(\rr_{\e\h})|\phi^\infty\rangle}{\langle\phi^\infty|\phi^\infty\rangle}\right\}\!,
  $$
where
  $$
\nabla^2V(\rr_{\e,\h})=V\!\left\{\frac2R\delta(r_{\e,\h}-R)+\delta'(r_{\e,\h}-R)\right\}\!,~~~~\nabla^2V_\C(\rr_{\e\h})=\frac{4\pi\alpha}{\varepsilon}\delta^{(3)}(\rr_{\e\h}),
  $$
and
  $$
\nabla^2W(\rr_{\e\h})=\frac W{R^2}\!\left\{6-8\frac{r_{\e\h}}{a^*}+4\frac{r^2_{\e\h}}{a^{*2}}\right\}\!\e^{-2\frac{r_{\e\h}}{a^*}}.
  $$
The previous mean values are indeed confinement regime dependent, as well as the Lamb shift, because of the IR cut-offs $\kappa^*_{\e,\h}$. Thus, we shall examine them in each confinement regime. To this end, we give some general results on diagonal matrix elements of the action of the Laplacian on potentials.

First, the case of Coulomb potential $V_\C(\rr_{\e\h})$ is trivially obtained since integrals over the electron (or the hole) coordinates are carried out over $\delta^{(3)}(\rr_{\e\h})$. This eliminates dependence on $\sigma_0$, so that
  $$
\langle\phi^\infty|\nabla^2V_\C(\rr_{\e\h})|\phi^\infty\rangle=\frac{e^2}\varepsilon\int\!\dd^3\rr\!~\psi^{\infty4}_{010}(\rr)=\frac{8e^2}{\kappa R^3}D
  $$
is independent from the confinement regime.

Following section {\bf\ref{sec_3}}, we have for the pseudo-potential $W(\rr_{\e\h})$
  $$
\langle\phi^\infty|\nabla^2W(\rr_{\e\h})|\phi^\infty\rangle=-W\frac8{R^4}\!\left\{6\pr_\Sigma-\frac8{a^*}\pr^2_\Sigma+\frac4{a^{*2}}\pr^3_\Sigma\right\}\!\frac1\Sigma\int_0^1\!\!\!\int_0^y\frac{\dd x}x\frac{\dd y}y\sin^2(\pi x)\sin^2(\pi y)\sinh(\Sigma Rx)\e^{-\Sigma Ry},
  $$
where $\Sigma=\sigma+\frac2{a^*}$ accounts for the exponential dependence of the pseudo-potential $W(\rr_{\e\h})$.

The determination of the formal quantities $\langle\phi^\infty|\nabla^2V^\infty(\rr_{\e,\h})|\phi^\infty\rangle$ is more cumbersome. As before, this is done using the regularization method of section {\bf\ref{sec_4}}, {\it i.e.} we expand the diagonal matrix element of the Laplacian of the finite potential step $\langle\phi|\nabla^2V(\rr_{\e,\h})|\phi\rangle$ in powers of $V$, and assume that $\langle\phi^\infty|\nabla^2V^\infty(\rr_{\e,\h})|\phi^\infty\rangle$ corresponds to the unique term independent from $V$. This calculation yields a term scaling as $\propto\sqrt V$, as in subsection {\it\ref{subsec_4_1}}. The results of appendixes {\it\ref{appendix_B_2}} and {\it\ref{appendix_B_3}} suggest to drop it, because it is not physical, so that
  $$
\langle\phi^\infty|\nabla^2V^\infty(\rr_{\e,\h})|\phi^\infty\rangle=\frac4{R^3}E^{\e,\h\infty}_{01}\pr_\sigma\!\left\{\e^{-\sigma R}\!\left[\frac3{\sigma R}-1\right]\!\int_0^1\!\frac{\dd x}x\sin^2(\pi x)\sinh(\sigma Rx)\right\}\!.
  $$
We now give the behavior of the Lamb shift of the confined electron-hole pair ground state in both confinement regimes for $CdS_{0.12}Se_{0.88}$ and $InAs$ quantum dots, with the numerical data collected in Table \ref{table_1}. The assumption of QD sphericity turns out to be accurate for QDs synthesized in such materials. In the strong confinement regime, we choose to consider a QD of radius of the order of magnitude of 10\AA~to adequately illustrate the properties of the predicted Lamb shift. There exists a large number of references on spherical $CdS_{0.12}Se_{0.88}$ or $InAs$ nanocrystals of radius of a few tens \AA, which are synthesized and experimentally used, {\it e.g.} see respectively \cite{Nomura_1990} or \cite{Banin_1998}.
\begin{table}
\caption{Numerical values of material parameters in $CdS_{0.12}Se_{0.88}$ and $InAs$, where the electron bare mass is $m_\e\approx9.11~\!10^{-31}$kg. $a_0\approx0.529$\AA~and $E_I\approx13.6$eV are the Bohr radius and the ionization energy of
the hydrogen atom.} \label{table_1}
  \begin{center}
{\footnotesize
    \begin{tabular}{cccc}
\hline
\hline
Semiconductor & in units of & $CdS_{0.12}Se_{0.88}$ & $InAs$
\\
\hline
$m^*_\e$ & $m_\e$ & 0.13 & 0.026
\\
$m^*_\h$ & $m_\e$ & 0.46 & 0.41
\\
$\varepsilon$ &  & 9.3 & 14.5
\\
$a^*$ & $a_0$ & 91.8 & 595
\\
$E^*$ & $E_I$ & 1.17~\!10$^{-3}$ & 1.16~\!10$^{-4}$
\\
\hline
\hline
    \end{tabular}}
  \end{center}
\end{table}
\subsection{Strong confinement regime}
Here $R\lesssim a^*$, Taylor expansions of the diagonal matrix elements of the action of the Laplacian on various potentials give the following expressions
  $$
  \left\{
    \begin{array}{rcl}
\ds\frac{\langle\phi^\infty|\nabla^2V^\infty(\rr_{\e,\h})|\phi^\infty\rangle}{\langle\phi^\infty|\phi^\infty\rangle}\lnd=\rnd\ds-\frac8{R^2}E^{\e,\h\infty}_{01}\!\left\{1-F\frac R{a^*}+F'\frac{R^2}{a^{*2}}+O\!\!\left(\frac{R^3}{a^{*3}}\right)\!\right\}\!,
\vs{.2cm}
\\
\ds\frac{\langle\phi^\infty|\nabla^2V_\C(\rr_{\e\h})|\phi^\infty\rangle}{\langle\phi^\infty|\phi^\infty\rangle}\lnd=\rnd\ds\frac{8e^2}{\kappa R^3}\!\left\{D+F''\frac R{a^*}+O\!\!\left(\frac{R^2}{a^{*2}}\right)\!\right\}\!,
\vs{.2cm}
\\
\ds\frac{\langle\phi^\infty|\nabla^2W(\rr_{\e\h})|\phi^\infty\rangle}{\langle\phi^\infty|\phi^\infty\rangle}\lnd=\rnd\ds-\frac{64}3\frac{E^\infty_{\e\h}}{a^{*2}}\!\left\{1+O\!\!\left(\frac R{a^*}\right)\!\right\}\!.
    \end{array}
  \right.
  $$
In the strong confinement regime, the electron and the hole are essentially uncorrelated, so that the IR cut-offs $\kappa^*_{\e,\h}$ should be chosen independently as, according to subsection {\it\ref{subsec_4_2}},
  $$
\kappa_{\e,\h}^*=\ds\frac{7\pi^2}{12m^*_{\e,\h}R^2}.
  $$
Omitting terms of the third order or higher in $\frac R{a^*}$ and dropping off the exponent ``~$^\infty$~'', we conclude that the Lamb shift undergone by the ground state of the electron-hole pair in the strong confinement regime is
  $$
\Delta E^\mathrm{strong}_\mathrm{Lamb}=\Delta E^\mathrm{strong}_\e+\Delta E^\mathrm{strong}_\h,
  $$
where
  $$
\frac{\Delta E^\mathrm{strong}_{\e,\h}}{E^\infty_{\e\h}}=-\frac{16\alpha}{3\pi\varepsilon}\frac{\lambdabarre^{*2}_{\e,\h}}{R^2}\log\!\!\left(\frac R{R^{\e,\h}_\mathrm{min}}\right)\!\!\left\{1-\!\left[\frac\mu{m^*_{\e,\h}}F+\frac2{\pi^2}D\right]\!\frac R{a^*}+\!\left[\frac\mu{m^*_{\e,\h}}F'-\frac2{\pi^2}F''+\frac83\right]\!\frac{R^2}{a^{*2}}\right\}\!\leq0,
  $$
$\ds\lambdabarre^*_{\e,\h}=m^{*-1}_{\e,\h}$ and $R^{\e,\h}_\mathrm{min}=\frac\pi2\sqrt{\frac73}\lambdabarre^*_{\e,\h}$ are respectively the reduced Compton wave lengths and minimal radii of the electron and of the hole in the considered semiconductor. At the zeroth order in $\frac R{a^*}$, the leading term to the Lamb shift of the electron-hole pair ground state is negative as a symptom of the {\it quasi}-uncorrelated confinement of the electron and the hole in the strong confinement regime. The first correction in $\frac R{a^*}$, particularly the contribution of the Coulomb potential $V_\C(\rr_{\e\h})$, is positive which recalls the behavior of the standard Lamb shift in real atoms.

  \begin{figure}
\caption{Lamb shift undergone by the electron (--$\!~$--), the hole (--$\!~\cdot~\!$--) and the exciton (---), when the exciton occupies its ground state in the strong confinement regime in {\bf a.} $CdS_{0.12}Se_{0.88}$ and {\bf b.} $InAs$ nanocrystals as a function of the QD radius.} \label{figure_1}
    \begin{center}
\begin{picture}(0,0)%
\includegraphics{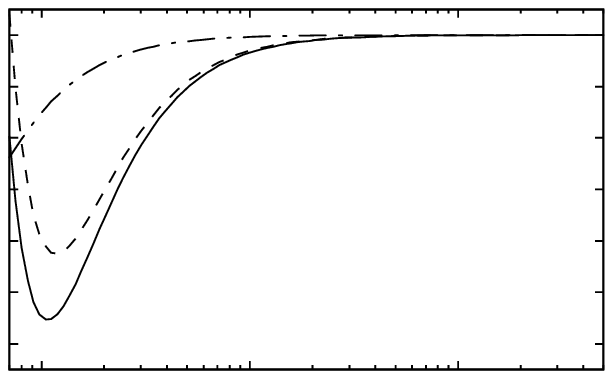}%
\end{picture}%
\setlength{\unitlength}{4144sp}%
\begingroup\makeatletter\ifx\SetFigFontNFSS\undefined%
\gdef\SetFigFontNFSS#1#2#3#4#5{%
  \reset@font\fontsize{#1}{#2pt}%
  \fontfamily{#3}\fontseries{#4}\fontshape{#5}%
  \selectfont}%
\fi\endgroup%
\begin{picture}(2974,1997)(-213,-990)
\put(2746,884){\makebox(0,0)[lb]{\smash{{\SetFigFontNFSS{10}{12.0}{\rmdefault}{\mddefault}{\updefault}{\color[rgb]{0,0,0}{\bf a.}}%
}}}}
\put(1383,-944){\makebox(0,0)[b]{\smash{{\SetFigFontNFSS{6}{7.2}{\rmdefault}{\mddefault}{\updefault}{\color[rgb]{0,0,0}$R$ (\AA)}%
}}}}
\put(-195, -5){\makebox(0,0)[rb]{\smash{{\SetFigFontNFSS{6}{7.2}{\rmdefault}{\mddefault}{\updefault}{\color[rgb]{0,0,0}$\ds\frac{\Delta E^{\mathrm{strong}}_{\mathrm{Lamb}}}{E^0_{\e\h}}$}%
}}}}
\put(  1,866){\makebox(0,0)[lb]{\smash{{\SetFigFontNFSS{5}{6.0}{\rmdefault}{\mddefault}{\updefault}{\color[rgb]{0,0,0}$\times10^{-4}$}%
}}}}
\put(-12,-719){\makebox(0,0)[rb]{\smash{{\SetFigFontNFSS{6}{7.2}{\rmdefault}{\mddefault}{\updefault}-0.6}}}}
\put(-12,-484){\makebox(0,0)[rb]{\smash{{\SetFigFontNFSS{6}{7.2}{\rmdefault}{\mddefault}{\updefault}-0.5}}}}
\put(-12,-248){\makebox(0,0)[rb]{\smash{{\SetFigFontNFSS{6}{7.2}{\rmdefault}{\mddefault}{\updefault}-0.4}}}}
\put(-12,-13){\makebox(0,0)[rb]{\smash{{\SetFigFontNFSS{6}{7.2}{\rmdefault}{\mddefault}{\updefault}-0.3}}}}
\put(-12,223){\makebox(0,0)[rb]{\smash{{\SetFigFontNFSS{6}{7.2}{\rmdefault}{\mddefault}{\updefault}-0.2}}}}
\put(-12,457){\makebox(0,0)[rb]{\smash{{\SetFigFontNFSS{6}{7.2}{\rmdefault}{\mddefault}{\updefault}-0.1}}}}
\put(-12,692){\makebox(0,0)[rb]{\smash{{\SetFigFontNFSS{6}{7.2}{\rmdefault}{\mddefault}{\updefault} 0}}}}
\put(174,-901){\makebox(0,0)[b]{\smash{{\SetFigFontNFSS{6}{7.2}{\rmdefault}{\mddefault}{\updefault} 0.1}}}}
\put(1126,-901){\makebox(0,0)[b]{\smash{{\SetFigFontNFSS{6}{7.2}{\rmdefault}{\mddefault}{\updefault} 1}}}}
\put(2078,-901){\makebox(0,0)[b]{\smash{{\SetFigFontNFSS{6}{7.2}{\rmdefault}{\mddefault}{\updefault} 10}}}}
\end{picture}~~~~~~~~~~~\begin{picture}(0,0)%
\includegraphics{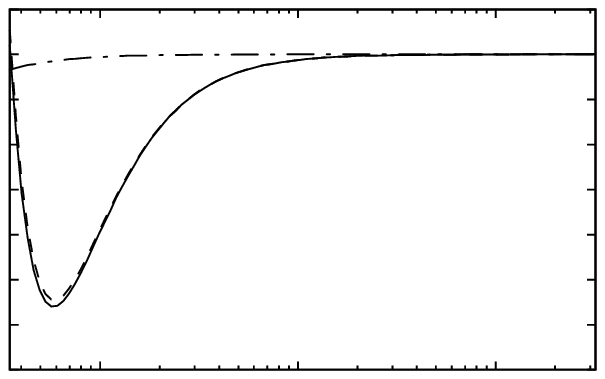}%
\end{picture}%
\setlength{\unitlength}{4144sp}%
\begingroup\makeatletter\ifx\SetFigFontNFSS\undefined%
\gdef\SetFigFontNFSS#1#2#3#4#5{%
  \reset@font\fontsize{#1}{#2pt}%
  \fontfamily{#3}\fontseries{#4}\fontshape{#5}%
  \selectfont}%
\fi\endgroup%
\begin{picture}(2935,2034)(-219,-1072)
\put(  1,848){\makebox(0,0)[lb]{\smash{{\SetFigFontNFSS{5}{6.0}{\rmdefault}{\mddefault}{\updefault}{\color[rgb]{0,0,0}$\times10^{-4}$}%
}}}}
\put(1363,-1026){\makebox(0,0)[b]{\smash{{\SetFigFontNFSS{6}{7.2}{\rmdefault}{\mddefault}{\updefault}{\color[rgb]{0,0,0}$R$ (\AA)}%
}}}}
\put(-195, -5){\makebox(0,0)[rb]{\smash{{\SetFigFontNFSS{6}{7.2}{\rmdefault}{\mddefault}{\updefault}{\color[rgb]{0,0,0}$\ds\frac{\Delta E^{\mathrm{strong}}_{\mathrm{Lamb}}}{E^0_{\e\h}}$}%
}}}}
\put(2701,839){\makebox(0,0)[lb]{\smash{{\SetFigFontNFSS{10}{12.0}{\rmdefault}{\mddefault}{\updefault}{\color[rgb]{0,0,0}{\bf b.}}%
}}}}
\put(-20,-651){\makebox(0,0)[rb]{\smash{{\SetFigFontNFSS{6}{7.2}{\rmdefault}{\mddefault}{\updefault}-0.3}}}}
\put(-20,-240){\makebox(0,0)[rb]{\smash{{\SetFigFontNFSS{6}{7.2}{\rmdefault}{\mddefault}{\updefault}-0.2}}}}
\put(-20,173){\makebox(0,0)[rb]{\smash{{\SetFigFontNFSS{6}{7.2}{\rmdefault}{\mddefault}{\updefault}-0.1}}}}
\put(-20,584){\makebox(0,0)[rb]{\smash{{\SetFigFontNFSS{6}{7.2}{\rmdefault}{\mddefault}{\updefault} 0}}}}
\put(430,-922){\makebox(0,0)[b]{\smash{{\SetFigFontNFSS{6}{7.2}{\rmdefault}{\mddefault}{\updefault} 1}}}}
\put(2239,-922){\makebox(0,0)[b]{\smash{{\SetFigFontNFSS{6}{7.2}{\rmdefault}{\mddefault}{\updefault} 100}}}}
\put(1335,-922){\makebox(0,0)[b]{\smash{{\SetFigFontNFSS{6}{7.2}{\rmdefault}{\mddefault}{\updefault} 10}}}}
\end{picture}%
    \end{center}
  \end{figure}

\begin{table}
\caption{Lamb shift undergone by the electron-hole pair ground state in $CdS_{0.12}Se_{0.88}$ or $InAs$ nanocrystals {\bf a.} for $R=10$\AA~or {\bf b.} for $R=30$\AA~in the strong confinement regime and {\bf c.} in the weak confinement regime.} \label{table_2}
  \begin{center}
{\footnotesize
    \begin{tabular}{cccc}
\hline
\hline
 & Semiconductor & $CdS_{0.12}Se_{0.88}$ & $InAs$
\\
\hline
{\bf a.} & $E^\infty_{\e\h}$ (eV) & 3.71 & 15.38
\vs{.1cm}
\\
 & $\frac{\Delta E^{\textrm{\tiny strong}}_\mathrm{Lamb}}{E^\infty_{\e\h}}$ & -5.52~\!10$^{-9}$ & -6.17~\!10$^{-7}$
\vs{.05cm}
\\
\hline
{\bf b.} & $E^\infty_{\e\h}$ (eV) & 0.412 & 1.71
\vs{.1cm}
\\
 & $\frac{\Delta E^{\textrm{\tiny strong}}_\mathrm{Lamb}}{E^\infty_{\e\h}}$ & -5.13~\!10$^{-9}$ & -8.64~\!10$^{-8}$
\vs{.05cm}
\\
\hline
\hline
    \end{tabular}
~~~~~~~~
    \begin{tabular}{cccc}
\hline
\hline
 & Semiconductor & $CdS_{0.12}Se_{0.88}$ & $InAs$
\\
\hline
{\bf c.} & $E^*$ (meV) & 15.9 & 1.58
\vs{.1cm}
\\
 & $\frac{\Delta E^\mathrm{weak}_\mathrm{Lamb}}{E^*}$ & 4.56~\!10$^{-7}$ & 1.70~\!10$^{-10}$
\vs{.05cm}
\\
\hline
\hline
\end{tabular}}
\end{center}
\end{table}

Figure \ref{figure_1} shows the behavior of the Lamb shift of the electron-hole pair ground state in spherical $CdS_{0.12}Se_{0.88}$ or $InAs$ QDs. Their order of magnitude suggests a possible detection for experimentally accessible QD radii. This is confirmed by table \ref{table_2} {\bf a.}-{\bf b.}, since the energy orders of magnitude are equivalent to those theoretically predicted and experimentally observed in hydrogen atom, at least in semiconductors with material parameters of the same order of magnitude than those of $InAs$, {\it e.g.} $GaAs$. Here, we do not identify the finite potential step $V(\rr_{\e\h})$ used for the regularization of $\langle\phi^\infty|V^\infty(\rr_{\e\h})|\phi^\infty\rangle$, which is a calculational intermediate tool, with the real potential step, which confines the charge carriers inside the QD and is a physical quantity of the problem. Results of sections {\bf\ref{sec_4}} and {\bf\ref{sec_5}} impose that electron and hole masses inside or outside the QD should be identical. This means that the charge carriers are first excited, and then confined. However, in reality, electrons and holes are excited in a semiconducting nanocrystal. In this situation, electrons and holes should have different masses inside as well as outside of the QD \cite{Thoai_1990}.

If we want to recognize the effective confining potential step at the surface of the nanocrystal with $V(\rr_{\e\h})$, then the possibility for the electron, and to a lesser extent the hole --- the notion of hole is not clearly defined in the surrounding insulating matrix ---, to leak out of the semiconducting QD by tunnel conductivity should be taken into account rigourously. Since sections {\bf\ref{sec_4}} and {\bf\ref{sec_5}} show that the only explicit dependence on the step potential appears as a non-physical term in the Lamb shift expression, this refinement is superfluous, reflecting the coherence of the reasoning.

Finally, in the strong confinement regime, the proposed analytical expression for the Lamb shift $\Delta E^\mathrm{strong}_\mathrm{Lamb}$ displays a linear dependence on the electron-hole pair confinement energy $E^\infty_{\e\h}$. This does not contradict the assumption that charge carriers are confined by a infinite potential well. Table \ref{table_2} shows that its order of magnitude exceeds the typical height of the effective potential step
at the surface of the QD. As shown by an analytical study of Stark effect in spherical semiconductor QDs \cite{Billaud_2009b}, the confinement energy itself should not be a relevant quantity, as far as energy shifts are concerned. Then, to validate the modeling of the confining potential by the infinite potential well, it is sufficient to compare the order of magnitude of Lamb shifts to the height of the real potential step. This is confirmed by table
\ref{table_2} {\bf a.}-{\bf b.}, since examples of Lamb shifts it presents are of the order of magnitude at most of tens $\mu$eV, a negligible value compared to the effective confinement potential, which is of the order of eV \cite{Thoai_1990}.
\subsection{Weak confinement regime} \label{subsec_7_3}
For $R\gtrsim\pi a^*$, Taylor expansions of the Laplacian action diagonal matrix elements yield
  $$
  \left\{
    \begin{array}{rcl}
\ds\frac{\langle\phi^\infty|\nabla^2V_\C(\rr_{\e,\h})|\phi^\infty\rangle}{\langle\phi^\infty|\phi^\infty\rangle}\lnd=\rnd\ds8\frac{E^*}{a^{*2}}\!\left\{1+\frac23\pi^2\frac{a^{*2}}{R^2}+O\!\!\left(\frac{a^{*3}}{R^3}\right)\!\right\}\!,
\vs{.2cm}
\\
\ds\frac{\langle\phi^\infty|\nabla^2W(\rr_{\e\h})|\phi^\infty\rangle}{\langle\phi^\infty|\phi^\infty\rangle}\lnd=\rnd\ds8\frac{E^*}{a^{*2}}\!\left\{-\frac23\pi^2\frac{a^{*2}}{R^2}+O\!\!\left(\frac{a^{*3}}{R^3}\right)\right\}\!.
    \end{array}
  \right.
  $$
In this weak confinement regime, the contribution of the pseudo-potential $W(\rr_{\e\h})$ to the electron-hole pair ground state Lamb shift cancels the second order correction of the contribution of the Coulomb potential $V_\C(\rr_{\e\h})$. Since they both scale as $\propto\pi^2\frac{E^*}{R^2}=\frac{E^\infty_{\e\h}}{a^{*2}}$, the presence of $W(\rr_{\e\h})$ allows the removal of contributions proportional to the electron-hole pair kinetic energy in relative coordinates, which are still contained in the exciton Rydberg energy. Finally, it is important to explain why it is not necessary to account for contributions coming from the confinement potential well $V^\infty(\rr_{\e\h})$ in the weak confinement regime. Direct calculations show that such contributions arise as a fifth order correction term in $\frac{a^*}R$ to $\frac{E^*}{a^{*2}}$
  $$
\frac{\langle\phi^\infty|\nabla^2V^\infty(\rr_{\e,\h})|\phi^\infty\rangle}{\langle\phi^\infty|\phi^\infty\rangle}\propto\frac{E^{0\infty}_{\e,\h}}{R^3}a^*\propto\frac{E^*}{a^{*2}}\frac{a^{*5}}{R^5}.
  $$
Omitting terms of the third order or higher in $\frac R{a^*}$ and dropping off the exponent ``~$^\infty$~'', we conclude that the Lamb shift undergone by the ground state of the electron-hole pair in the weak confinement regime is
  $$
\Delta E^\mathrm{weak}_\mathrm{Lamb}=\Delta E^\mathrm{weak}_\e+\Delta E^\mathrm{weak}_\h,\textrm{~~~~where~~~~}\frac{\Delta E^\mathrm{weak}_{\e,\h}}{E^*}=\frac{8\alpha}{3\pi\varepsilon}\frac{\lambdabarre^{*2}_{\e,\h}}{a^{*2}}\log\!\!\left(\frac{m_{\e,\h}^*}{\kappa^*_{\e,\h}}\right)\!.
  $$
In the limit of infinite hole mass, {\it i.e.} in the limit $\frac{\lambdabarre^*_\h}{\lambdabarre^*_\e}=\frac{m^*_\e}{m^*_\h}\rightarrow0$, in the weak confinement regime, only the electronic term $\Delta E^\mathrm{weak}_\e$ contributes to the exciton ground state Lamb shift. Then, after some trivial manipulations, the Lamb shift of the ground state of an hydrogen-like atom of reduced mass $\mu$ in a dielectric medium characterized by its dielectric constant $\varepsilon$ is recognized. These observations show that electron-hole pair states behave as excitonic bound states in this regime. Using known results for the hydrogen-like atom \cite{Bethe_1950}, they suggest to take an acceptable approximate value for the IR cut-off as
  $$
\kappa^*_{\e,\h}\approx19.8E^*.
  $$
Thus, the Lamb shift undergone by the exciton ground state does not depend on the QD radius, at least up to the third order in $\frac{a^*}R$. This is a strong argument in favor of the validity of the pseudo-potential $W(\rr_{\e\h})$, which was introduced as a phenomenological effective potential.

Table \ref{table_2} {\bf c.} gives values of the exciton ground state Lamb shift in $CdS_{0.12}Se_{0.88}$ or $InAs$ nanocrystals. They are not experimentally accessible at present.
\subsection{Observability of the Lamb shift in spherical semiconductor QDs}
As known, the experimental observability of the Lamb shift in hydrogen atom is due to the $s$- and $p$-level degeneracy, when the principal quantum number is $n\geq 2$, in absence of interaction with the quantized electromagnetic field. The Lamb shift arises as a separation of the $ns$-spectral band from $np$-spectral band, while they should stay merged in absence of Lamb effect. So, how would an energy level Lamb shift be detected for quantum systems displaying no spectral band degeneracy such as a QD? In such systems each non-degenerate energy level is dressed by the quantum zero-point fluctuations of the electromagnetic field, forbidding the detection of the corresponding bare level. This may be the reason for the lack of articles addressing the Lamb effect in such structures.

In Quantum Field Theory, the summation of the zero-point energy fluctuations yields a divergent ground state energy. In absence of gravity, this divergence is subtracted off in an additive renormalization scheme. However, a careful analysis on its volume dependence (via boundary conditions) shows the occurrence of a finite and observable force, known as Casimir force \cite{Casimir_1948, Mohideen_1998}. This effect can be intuitively understood as follows. In vacuum, two parallel perfectly conducting squared plates of linear size $L$ are placed at a separation $d\ll L$. Since the zero-point energy fluctuations are more important outside than inside the plates, they are subjected to an attractive the Casimir force.

If we adopt the Welton framework, we may view the Lamb shift as due to the particle position fluctuations induced by the zero-point energy fluctuations energy of the electromagnetic field. So, by placing a QD in two different quantized electromagnetic surroundings (in vacuum and inside a Casimir pair of conducting plates), one would be able to detect an energy level difference between two Lamb shifted levels. An experimental protocol according to which the energy levels dressed by the zero-point fluctuations energy with or without the two Casimir plates are to be compared. It should allow
to overcome the need of degenerate energy levels, or of exactly
computed energy levels.

There exist some theoretical works dealing with Lamb effect of real atoms confining in a Casimir device, but they are based on a relativistic Bethe approach \cite{Cheon_1988, Jhe_1991}. They predict an additional shift to the standard Lamb shift, which depends on the separation distance between the mirrors such that it goes to zero in the limit $d\rightarrow\infty$. On the other hand the Welton approach suggests, in this limit, that the Lamb shift in presence of the Casimir device should go to the Lamb shift in absence of such device. While the predicted additional shifts should be experimentally measured, they seem to not be available at present in the literature. The protocol suggested here is meant to reinforce the complementarity of Bethe and Welton approaches. This is the reason why, as a continuation of this work, we shall investigate a generalization of the Welton approach in order to evaluate the particle position fluctuations, in the presence of two Casimir plates.
\section{Conclusion}
Following the conventional approaches for computing the Lamb shift in real atoms, we have worked out an analytical expression of the Lamb shift for the ground state of a electron-hole-pair confined in a spherical semiconductor QD. An explicit expression is obtained in the framework of the EMA model augmented by the Coulomb interaction, for both strong and weak confinement regimes. We find that, in the strong confinement regime and for sufficiently small but experimentally realistic QD sizes, this Lamb shift has an order of magnitude comparable to the Lamb shift in atoms, but is of opposite sign.

The observability of such a Lamb shift is put to question since the energy levels are not degenerate as in the hydrogen atom. A way out, based on the Welton approach, is suggested. It consists in comparing the energy levels of the same spherical semiconductor QD in two different quantized electromagnetic field environments: the vacuum and a Casimir cavity. It is amusing to see that the two foremost effects which have served to validate QED over the years may be now reunited to demonstrate the existence of the Lamb shift in semiconductor QDs.

These theoretical considerations are to be confirmed by experimental work. If the orders of magnitude predicted here are confirmed for the ground state of a confined electron-hole pair in the strong confinement regime, it would be worth while to generalize the method we have presented to any energy level of the pair. Therefore, the issue of how to account for the Lamb effect in modern numerical methods, such as DFT or {\it ab initio} methods, should be put in question. However, advances in such numerical methods are to be developed first. At the moment, they do not actually allow to compute energy levels of semiconducting nanostructures with sufficient accuracy \cite{Delerue, Bester_2009} to observe effects of orders of magnitude of the Lamb shift.

Finally, among several theories improving the description of semiconducting QDs by EMA models and their theoretical predictions \cite{Yoffe_2002}, let us mention a sophisticated version of the EMA, build on a reformulation of the multiband envelop theory \cite{Bastard_1986}. This is an extension of the $\kk\cdot\pp$ perturbation theory, which includes non-parabolicity in the calculation of band structure in semiconducting heterostructures \cite{Baldereschi_1973, Baldereschi_1974}. In the spherical approximation, the band structure is analytically determined in a neighborhood of a point of the Brillouin zone, typically its center, the so-called $\Gamma$-point, by second order development \cite{Sercel_1990, Vahala_1990}. It is natural to ask if it is possible to extend the theoretical approach presented in this paper to study the Lamb shift in spherical semiconducting QDs, with a standard EMA model, to the case of a multiband envelop theory, since all needed essential mathematical tools already exist. In particular, a orthonormal basis of eigen-functions of the QD Hamiltonian is explicitly known. The difficulty would therefore lie in the exact computation of the Laplacian matrix elements. However the feasibility of this ask does not seem, at least for the moment, to be attainable.
\appendix
\section{Constants} \label{appendix_A}
In the following tables, we sum up all appearing constants and give their approximate values. The function $\Si(x)=\int_0^x\frac{\dd t}t\sin(t)$ denotes  the standard sine integral. Table \ref{table_4} {\bf a.} (resp. Table \ref{table_4} {\bf b.}) gives the analytical expressions and approximate values of constants occurring in section {\bf \ref{sec_3}} (resp. in section {\bf\ref{sec_7}}).

\begin{table*}
\caption{Definitions, analytic expressions and approximate values of constants appearing {\bf a.} in section {\bf \ref{sec_3}}, and {\bf b.} in section {\bf \ref{sec_7}}, where $\sigma'_0=\sigma_0a^*$ is the dimensionless variational parameter in the strong confinement regime.} \label{table_4}
\begin{center}
{\footnotesize
\begin{tabular}{ccccccc}
\hline
\hline
 & Name & Expression & Value & Name & Expression & Value
\\
\hline
{\bf a.} & $S$ & $\Si(2\pi)-\frac{\Si(4\pi)}2$ & 0.6720 & $A$ & $2-\frac S\pi$ & 1.7861
\vs{.1cm}
\\
 & $B_1$ & $\frac23-\frac5{8\pi^2}$ & 0.6033 & $B_2$ & $\frac29+\frac{13}{24\pi^2}+\frac S{2\pi^3}$ & 0.2879
\vs{.1cm}
\\
 & $B$ & $B_1+\frac{B_2}3$ & 0.6993 & $B'$ & $AB-1$ & 0.2489
\vs{.1cm}
\\
 & $C$ & $\frac13-\frac1{2\pi^2}$ & 0.2827
\vs{.1cm}
\\
\hline
{\bf b.} & $\sigma'_0$ & $4B'$ & 0.9956 & $D$ & $\frac\pi2S$ & 1.0557
\vs{.1cm}
\\
 & $F$ & $\left\{\frac{C+5}4-B\right\}\!\sigma'_0$ & 0.6187 & $F'$ & $\left\{B^2-B\frac{C+5}4-\frac{2C-3}4\right\}\!\sigma'^2_0$ & 0.6993
\vs{.1cm}
\\
 & $F''$ & $BD\sigma'_0$ & 0.7350
\vs{.1cm}
\\
\hline
\hline
\end{tabular}}
\end{center}
\end{table*}
\section{Bethe approach to the Lamb shift of a particle confined in a spherical potential well} \label{appendix_B}
The wave functions of the confined particle confined by the spherical finite potential step $V(\rr)$ are defined, for quantum numbers $l\in\mathbb N$, $n\in\mathbb N\smallsetminus\{0\}$ and $m\in\ldc-l,l\rdc$, as
  \begin{eqnarray*}
\psi_{lnm}(\rr)\lnd=\rnd\!\left\{R_{ln}^\I(r)\chi_{[0,R[}(r)+R_{ln}^{\I\I}(r)\chi_{[R,\infty[}(r)\right\}\!Y^m_l(\theta,\varphi)
\\
\lnd=\rnd\left\{A_{ln}\j_l\!\!\left(\frac{k_{ln}}Rr\right)\!\chi_{[0,R[}(r)+C_{ln}\h^{(1)}_l\!\!\left(i\frac{K_{ln}}Rr\right)\!\chi_{[R,\infty[}(r)\right\}\!Y^m_l(\theta,\varphi).
  \end{eqnarray*}
Because the limit $V\rightarrow\infty$ is of interest, it is possible to adjust the step potential height $V$, so that the energy eigenvalue $E_{ln}$ is written as functions of the wave numbers $k_{ln}$ and $K_{ln}$ in regions I and II
  $$
E_{ln}=\frac{k_{ln}^2}{2m^*R^2}\textrm{~~~~and~~~~}V-E_{ln}=\frac{K^2_{ln}}{2m^*R^2}\geq0.
  $$
The normalization coefficients $A_{ln}$ and $C_{ln}$, and the wave numbers $k_{ln}$ and $K_{ln}$ are subjected to boundary conditions at $r=R$, {\it i.e.}
  $$
    \left\{
    \begin{array}{rcl}
A_{ln}\j_l(k_{ln})\lnd=\rnd C_{ln}\h^{(1)}_l(iK_{ln})
\vs{.1cm}
\\
A_{ln}k_{ln}\j_{l\pm1}(k_{ln})\lnd=\rnd iC_{ln}K_{ln}\h_{l\pm1}^{(1)}(iK_{ln})
      \end{array}
    \right.
  $$
and by orthonormalization conditions. In region I, the radial wave function $R_{ln}^\I(r)$ has the same form as the radial wave function $R_{ln}^{\I\infty}(r)$, which is then really the limit of $R_{ln}^\I(r)$, when $V\rightarrow\infty$. The same remark should be made for the radial wave function $R_{ln}^{\I\I}(r)\xrightarrow[V\infty]{}0$, since, by construction, $K_n\xrightarrow[V\infty]{}\infty$. Nevertheless, the introduction of the potential step allows the particle to get out the confining potential by tunnel effect, even if the probability for such event is exponentially small.
\subsection{Lamb shift via the Poisson Law} \label{appendix_B_1}
The potential step $V(\rr)$ satisfies the Poisson equation $\nabla^2V(\rr)=V\!\left\{\frac2R\delta(r-R)+\delta'(r-R)\right\}$, $\delta(r)$ being the Dirac distribution and $\delta'(r)$ its derivative. In the regularization method introduced in section {\bf\ref{sec_4}}, the unique term independent of $V$ of the diagonal matrix element of the Laplacian of the potential step $V(\rr)$ expansion in powers of $V$ should be retrieved when $V\rightarrow\infty$, since it is the only non-divergent and non-vanishing one. This expansion in powers of $V$ can be explicitly done, and yields
  $$
\langle\psi_{lnm}|\nabla^2 V(\rr)|\psi_{lnm}\rangle=-\frac{4V}\pi |A_{ln}|^2 k^2_{ln}\j_l(k_{ln})\j_l'(k_{ln})=\frac2{R^3}\sqrt{\frac{2V}{m^*}}(k_{ln}^\infty)^2-\frac8{R^2}E_{ln}^\infty+o(1).
  $$
This suggests that the Lamb shift undergone by the particle confined by the infinite potential well $V^\infty(\rr)$ is actually given by Eq. (\ref{DeltaE^infty_lamb}).
\subsection{Alternative route to the Lamb shift in the Coulomb gauge} \label{appendix_B_2}
An alternative calculation of the Lamb shift undergone by a particle confined in the infinite potential well $V^\infty(\rr)$ is presented here. Starting with the regularized version by the finite potential step $V(\rr)$ of Eq. (\ref{DeltaE_lamb}), but for which the closure relation is not used
  $$
\Delta E_{lnm}=\frac2{3\pi}\frac{\alpha^2}{m^{*2}}\log\!\!\left(\frac{m^*}\kappa\right)\sum_{ijk} \left|\langle\psi_{ijk}|\pp|\psi_{lnm}\rangle\right|^2\{E_{ij}-E_{ln}\},
  $$
we directly evaluate the sum $\sum_{ijk}\left|\langle\psi_{ijk}|\pp|\psi_{lnm}\rangle\right|^2\{E_{ij}-E_{ln}\}=-\langle\psi_{lnm}|\nabla\cdot H_0\nabla-E_{ln}\nabla^2|\psi_{lnm}\rangle$, and expand it in powers of $V$, as already mentioned in subsection {\it \ref{subsec_4_1}}. The obtained expansion does not have divergent terms when $V\rightarrow\infty$, such as terms scaling as $\propto\sqrt V$ of appendix {\it \ref{appendix_B_1}}. So, obtaining the Lamb shift for a state confined by the infinite potential well from the one confined by the finite potential step defined by the same quantum numbers is rigorously justified.

Actually, it can be shown that
  $$
\left\{\nabla\cdot H_0\nabla-E_{ln}\right\}\psi_{lnm}(\rr)=\frac{R_{ln}^{\I\prime\prime}(R)-R_{ln}^{\I\I\prime\prime}(R)}{2m^*}\!\left\{\delta'(r-R)+\frac2R\delta(r-R)\right\}\!Y^m_l(\theta,\varphi),
  $$
from which one gets
  $$
\sum_{ijk}\left|\langle\psi_{ijk}|\pp|\psi_{lnm}\rangle\right|^2\{E_{ij}-E_{ln}\}=\frac{R^2}{2m^*}R_{ln}^{\I\infty\prime\prime}(R)R_{ln}^{\I\infty\prime}(R)+o(1)=-4\frac{E^\infty_{ln}}{R^2}+o(1).
  $$
As stated, the first term in this expansion, and therefore also in the Lamb shift expansion, is the finite term, which leads to a Lamb shift  identical to Eq. (\ref{DeltaE^infty_lamb}).
\subsection{Alternative route to the Lamb shift in the electric dipole approximation} \label{appendix_B_3}
Applying the reasoning of appendix {\it \ref{appendix_B_2}} to the dipole approximation on  Eq. (\ref{DeltaE_lamb'}), we obtain once again the Lamb shift given by Eq. (\ref{DeltaE^infty_lamb}). Using expressions
  \begin{eqnarray*}
\Delta E_{lnm}\lnd=\rnd\frac{2\alpha}{3\pi}\log\!\!\left(\frac{m^*}\kappa\right)\!\sum_{ijk}\left|\langle\psi_{ijk}|\rr|\psi_{lnm}\rangle\right|^2(E_{ij}-E_{ln})^3
\\
\lnd=\rnd\frac{2\alpha}{3\pi}\log\!\!\left(\frac{m^*}\kappa\right)\!\langle\psi_{lnm}|[H_0,\rr]^\dag\!\cdot\!\left[H_0,[H_0,\rr]\right]\!|\psi_{lnm}\rangle,
  \end{eqnarray*}
and computing the following commutators
  $$
[H_0,\rr]\psi_{lnm}(\rr)=-\frac1{m^*}\!\left\{\pr_r\rr\pr_r+\pr_\theta\rr\frac{\pr_\theta}{r^2}+\pr_\varphi\rr\frac{\pr_\varphi}{r^2\sin^2\theta}\right\}\!\psi_{lnm}(r,\theta,\varphi),
  $$
and
  $$
\left[H_0,[H_0,\rr]\right]\psi_{lnm}(\rr)=-\frac{R_{ln}^{\I\prime\prime}(R)-R_{ln}^{\I\I\prime\prime}(R)}{2m^{*2}}\delta(r-R)\frac\rr R,
  $$
we come up with
  \begin{equation*}
\langle\psi_{lnm}|[H_0,\rr]^\dag.\left[H_0,[H_0,\rr]\right]\!|\psi_{lnm}\rangle=-\frac4{m^{*2}}\frac{E^\infty_{ln}}{R^2}+o(1).
  \end{equation*}
Hence, we can conclude that this Lamb shift is identical to the one given by Eq. (\ref{DeltaE^infty_lamb}). Let us also note that this method does not generate divergent terms in the limit $V\rightarrow\infty$, as in appendix {\it\ref{appendix_B_2}}.
\section{Feynman regularization of the infrared cut-off $\kappa$} \label{appendix_C}
To determine an accurate value for the IR cut-off $\kappa$, the Feynman regularization method is used instead of the regularization method based on the existence of the UV cut-off $\kappa_\mathrm{lim}$ for authorized wave numbers mentioned in subsection {\it\ref{subsec_4_2}}. Beginning with Eq. (\ref{kappa}), a dimensionless parameter $\epsilon>0$ is introduced, and the regularized sum
  $$
\kappa=\frac1{2m^*R^2}\frac{\sum_{ijln}(2i+1)(2l+1)|(k^\infty_{ln})^2-(k^\infty_{ij})^2|\e^{-i\epsilon}\e^{-j\epsilon}\e^{-l\epsilon}\e^{-n\epsilon}}{\sum_{ijln}(2i+1)(2l+1)\e^{-i\epsilon}\e^{-j\epsilon}\e^{-l\epsilon}\e^{-n\epsilon}}.
  $$
should be expanded in powers of $\epsilon$. The finite part of this expansion should be kept and identified with the IR cut-off $\kappa$. Considering the Bessel functions asymptotical behavior and their roots repartition at infinity \cite{Gradshteyn}, for sufficiently large quantum number $q\in\mathbb N\smallsetminus\{0\}$, but for all $p\in\mathbb N$, we can assume that
  $$
k^\infty_{pq}\approx k^\infty_{p1}+(q-1)\pi\approx k^\infty_{01}+(q-1)=q\pi.
  $$
While this approximation is not valid for any quantum numbers $q\in\mathbb N$, in the limit $\epsilon\rightarrow0$, the most important contribution to $\kappa$ corresponds to high quantum numbers, {\it i.e.} it should be used for all wave numbers. Then,
  $$
\kappa=\frac{\pi^2}{2m^*R^2}\frac{\sum_{jn}|n^2-j^2|\e^{-j\epsilon}\e^{-n\epsilon}}{\sum_{jn}\e^{-j\epsilon}\e^{-n\epsilon}}=\frac{\pi^2}{2m^*R^2}\left\{\frac4{\epsilon^2}+\frac3{\epsilon}+\frac76+o(1)\right\}\!,
  $$
which yields to the definition for $\kappa$, given in subsection
{\it\ref{subsec_4_2}}.

\end{document}